# Bond-selective modulation of scalar spin chirality in strained $Mn_4N$

Xin Liu, Mingyue Zhao, Shun Niu, Yu Liu, Yang Li, Jiayao Zhu, Yiwen Zhang, Fengxian Ma, Dewei Zhao, Guoke Li, Li Ma*, Congmian Zhen, Denglu Hou

*Hebei Key Laboratory of Photophysics Research and Application, College of Physics, Hebei Normal University, Shijiazhuang, 050024, China.*

**Abstract**

Scalar spin chirality (SSC) underlies a variety of Berry-phase-driven transport phenomena in noncoplanar magnets. $Mn_4N$ is a unique ferrimagnetic system in which both collinear and noncoplanar magnetic configurations have been reported at different lattice parameters, featuring vanishing and finite SSC, respectively. However, the microscopic mechanism governing the competition between these magnetic states and the associated modulation of SSC remains unclear. This issue is particularly important for $Mn_4N$ because its high magnetic ordering temperature ($T_N$ = 740 K) provides an attractive platform for exploring chiral magnetic states and their associated topological functionalities at elevated temperatures. Here, using first-principles calculations, we investigate the strain-driven evolution of the magnetic ground state and SSC in $Mn_4N$ and uncover the microscopic origin of the collinear-to-noncoplanar magnetic transition. We demonstrate that tensile strain continuously stabilizes the noncoplanar configuration and enhances SSC, as quantified by the magnitude of the chirality-order vector. Orbital-resolved crystal orbital Hamilton population and charge-density-difference analyses reveal that strain selectively weakens the Mn3c-N hybridization while preserving direct Mn3c-Mn3c interactions, leading simultaneously to the activation of the Mn3c in-plane magnetic moment and the suppression of N-mediated ferromagnetic superexchange between nearest-neighbor Mn3c atoms. This bond-selective electronic response governs the magnetic ground state and consequently controls the emergence and enhancement of SSC in $Mn_4N$. Our work establishes bond selectivity as an effective strategy for engineering SSC in high-temperature ferrimagnetic materials.

* Corresponding author. E-mail address: majimei@hebtu.edu.cn

**Introduction**

A series of novel physical phenomena in noncoplanar magnets has been observed, including the topological Hall effect[1-4], anomalous Hall effect[5-8], topological Nernst effect[9,10], thermal Hall effect[11,12], higher-order harmonics of the planar Hall effect[13], spin splitting[14], and topological magneto-optical Kerr effect[15,16]. To quantitatively characterize the degree of noncoplanarity in a triangular plaquette, scalar spin chirality (SSC) $\chi_{ijk}$ with scalar properties is introduced and defined as $\chi_{ijk} = \boldsymbol{S}_i \cdot (\boldsymbol{S}_j \times \boldsymbol{S}_k)$, where $\boldsymbol{S}_\alpha\ (\alpha = i, j, k)$ represents the spins of three magnetic atoms forming the triangular plaquette.[3,17-19] The origins of these physical phenomena can be mainly attributed to the following mechanisms: intrinsic contributions of SSC[1,3,5], SSC fluctuations[12,20,21], and SSC induced skew scattering[11,22,23]. Theoretical studies further predict that the intrinsic SSC would also induce phenomena such as the quantum anomalous Hall effect[24-27], quantum topological Hall effect[26,28], and quantum topological magneto-optical Kerr effect[29].

In this context, we have summarized the materials hosting intrinsic SSC and listed them in Table 1. It is evident that intrinsic SSC systems are relatively rare, which may be due, to some extent, to SSC cancellation by lattice symmetry.[17] Notably, $Mn_4N$ stands out with a high Néel temperature of approximately 740 K[30], making it a highly promising candidate for high-temperature topological spintronic devices. However, the magnetic ground state of $Mn_4N$ remains a complex and long-standing issue that has been actively debated.[31,32] At first, a collinear ferrimagnetic configuration in $Mn_4N$ was confirmed through neutron diffraction.[33,34] Based on this collinear model, experimental[35-40] and theoretical[40-42] investigations have been conducted to date. Subsequently, Fruchart *et al.* performed the polarized neutron diffraction experiments and found a finite noncollinear component in the (111) plane of $Mn_4N$.[30,43] Mohn *et al.* proposed the noncoplanar magnetic structure by ab initio analysis.[44] Isogami *et al.* also found noncoplanar magnetic structures in $Mn_4N$ epitaxial films.[45] Recently, based on this noncoplanar model, excellent performance in $Mn_4N$ has been both theoretically predicted[46,47] and experimentally reported[47,48]. Interestingly, He *et al.* [46] and Zemen[49] predicted that both collinear and noncoplanar models can be stabilized in $Mn_4N$ depending on the lattice constant. It is worth noting that the collinear and noncoplanar ferrimagnetic states correspond to zero and finite SSC, respectively. Therefore, elucidating the underlying microscopic physical mechanism governing the magnetic ground state of $Mn_4N$ as a function of lattice constant is crucial for understanding the emergence of SSC.

Table 1. List of materials hosting scalar spin chirality. We group the materials into two archetypes based on the scalar spin chirality origin: inherent and induced. We also list the induced approach of scalar spin chirality. Next, we list the magnetic materials platform and temperature of the experiments ($T$). Finally, we list the magnetic lattice

type.

| SSC Origin | Induced Approach | Materials | *T* (K) | Magnetic Lattice |
|---|---|---|---|---|
| inherent | — | $CoNb_3S_6$[3] | ~28 | triangular |
| | | $CoTa_3S_6$[3] | ~24 | triangular |
| | | $MnTe_2$[14] | ~86.5 | triangular |
| | | $UCu_5$[6] | ~1.2 | triangular |
| | | $Nd_2Mo_2O_7$[1] | ~40 | pyrochlore |
| | | $Pr_2Ir_2O_7$[5,19] | ~40 | pyrochlore |
| | | $Nd_2Ir_2O_7$[13] | ~30 | pyrochlore |
| | | $Mn_3Sn$[50] | ~50 | kagome |
| | | $Mn_4N$[30] | ~740 | triangular |
| induced | chemical doping | $Mn_{3-\delta}Fe_{\delta}Sn$[51] | ~300 | kagome |
| | | $Fe_{1+\delta}Sb$[18] | ~30 | triangular |
| | | $Mn_{3+\delta}Ni_{1-\delta}N$[52] | ~228 | kagome |
| | | $(Y_{1-x-y}Tb_xCd_y)_2Mo_2O_7$[53] | ~20 | pyrochlore |
| | magnetic field | $PdCrO_2$[2] | ~20 | triangular |
| | | $Mn_3Sn$[54] | ~350 | kagome |
| | | $Mn_3Pt$[7] | ~340 | kagome |
| | pressure | $Mn_3Ge$[55] | ~300 | kagome |

In this work, the evolution of the collinear and noncoplanar magnetic ground states in strained $Mn_4N$ is investigated using first-principles calculations. We find that the degree of noncoplanarity in a magnetic tetrahedron, quantified by the magnitude of the chirality-order vector $|\chi|$, can be continuously tuned by isotropic strain. Charge density difference and projected crystal orbital Hamilton populations analyses reveal that the selective weakening of Mn–N covalent bonds, while largely preserving direct Mn–Mn interactions, provides the microscopic mechanism governing the strain-induced modulation of scalar spin chirality.

**Calculation Method**

All first-principles calculations were performed within the framework of density functional theory (DFT) utilizing the Vienna Ab Initio Simulation Package (VASP)[56-58]. The interactions between core electrons and valence ions were described using the projector-augmented wave (PAW) method. For the exchange-correlation functional, the generalized gradient approximation (GGA) parameterized by the Perdew-Burke-Ernzerhof (PBE) formalism was adopted. A plane-wave energy cutoff of 520 eV and a $\Gamma$-centered 12×12×12 $k$-point mesh for Brillouin zone sampling were used. The convergence criteria for the electronic self-consistent field and ionic relaxation were strictly set to $10^{-6}$ eV/cell and $10^{-3}$ eV/Å, respectively. The isotropic strain calculations were performed by applying a uniform expansion or compression to the cubic $Mn_4N$ lattice, in which all three lattice constants were scaled simultaneously by the same strain ratio while preserving the cubic symmetry. At each strain value, the internal atomic coordinates were fully relaxed until the residual forces satisfied the convergence criterion $10^{-3}$ eV/Å. The same structural relaxation procedure was employed for both the collinear and noncoplanar magnetic configurations to ensure a consistent comparison of their total energies. Besides, additional biaxial strain calculations were performed to simulate experimentally relevant epitaxial $Mn_4N$ thin films grown on $SrTiO_3$ (111) and MgO (111) substrates. Crucially, spin-orbit coupling (SOC) was explicitly incorporated in all static electronic structure calculations based on the fully relaxed crystal structures.

To extract the magnetic exchange interactions, the energy mapping four-states method[59-61] was employed. In this method, the exchange interaction constant $J_{12}$ between the selected magnetic pair is calculated using the expression:

$$E_{\text{spin}} = J_{12}\boldsymbol{S}_1 \cdot \boldsymbol{S}_2 + \boldsymbol{S}_1 \cdot \boldsymbol{K}_1 + \boldsymbol{S}_2 \cdot \boldsymbol{K}_2 + E_{\text{other}} \tag{1}$$

Where $\boldsymbol{S}_1$ and $\boldsymbol{S}_2$ represent the spins of the selected magnetic pair, respectively, $\boldsymbol{K}_1 = \sum_{i \neq 1,2} J_{1i}\boldsymbol{S}_i$, $\boldsymbol{K}_2 = \sum_{i \neq 1,2} J_{2i}\boldsymbol{S}_i$, and $E_{\text{other}} = \sum_{i,j \neq 1,2} J_{ij}\boldsymbol{S}_i \cdot \boldsymbol{S}_j$. The first term is the exchange interaction between the selected magnetic pair, the second (third) term is the exchange interaction between $\boldsymbol{S}_1$ ($\boldsymbol{S}_2$) and other magnetic atoms $\boldsymbol{S}_i$ ($\boldsymbol{S}_j$), and the fourth term is the exchange interaction between other magnetic atoms. Four collinear spin states are considered: (1) $\boldsymbol{S}_1 = S$, $\boldsymbol{S}_2 = S$, (2) $\boldsymbol{S}_1 = S$, $\boldsymbol{S}_2 = -S$, (3) $\boldsymbol{S}_1 = -S$, $\boldsymbol{S}_2 = S$, (4) $\boldsymbol{S}_1 = -S$, $\boldsymbol{S}_2 = -S$. These four states have the following energy expressions:

$$\begin{aligned} E_1 &= E_0 + E_{\text{other}} + J_{12}S^2 + K_1S + K_2S \\ E_2 &= E_0 + E_{\text{other}} - J_{12}S^2 + K_1S - K_2S \\ E_3 &= E_0 + E_{\text{other}} - J_{12}S^2 - K_1S + K_2S \\ E_4 &= E_0 + E_{\text{other}} + J_{12}S^2 - K_1S - K_2S \end{aligned} \tag{2}$$

Then $J_{12}$ can be calculated as:

$$J_{12} = \frac{E_1 + E_4 - E_2 - E_3}{4S_1S_2} \tag{3}$$

From this, the exchange coupling constant $J_{12}$, which governs the interatomic magnetic interactions, can be obtained. In addition, the spin model proposed by Szunyogh *et al.* in Ref. [62] also provides an approach to obtain the exchange interaction constant. Both approaches describe the microscopic interactions of magnetic systems by mapping them onto a classical Heisenberg-type spin model, with parameters derived from first-principles DFT calculations. In our work, we only used the four-states method to calculate the exchange coupling constant $J$ between the nearest-neighbor Mn3c atoms. Therefore, both $\boldsymbol{S}_1$ and $\boldsymbol{S}_2$ in Eq. (1) correspond to Mn3c spins.

Following Refs. [1,18], the chirality associated with each triangular plaquette is oriented along its normal direction. We therefore define the chirality-order vector $\boldsymbol{\chi}$ of each magnetic tetrahedron in Fig. 1(b) as

$$\boldsymbol{\chi} = \sum_{\langle ijk \rangle} \boldsymbol{S}_i \cdot (\boldsymbol{S}_j \times \boldsymbol{S}_k) \boldsymbol{n}_{ijk} \tag{4}$$

where $\boldsymbol{S}_i \cdot (\boldsymbol{S}_j \times \boldsymbol{S}_k)$ denotes the scalar spin chirality of the triangular plaquette formed by the spins at sites *i*, *j*, and *k*, and $\boldsymbol{n}_{ijk}$ is the unit normal vector of the corresponding plaquette. The summation runs over the four triangular plaquettes of the tetrahedron. The resulting $\boldsymbol{\chi}$ characterizes the collective chirality configuration of the tetrahedron. By comparison, the vector chirality $\boldsymbol{\kappa}$ is calculated as the sum of the vector spin chirality between spin pairs in a triangular plaquette[62], which characterizes the noncollinearity of spins in the triangular plaquette. From this, it is evident that although both $\boldsymbol{\chi}$ and $\boldsymbol{\kappa}$ are vectors, they originate from different spin chiralities.

We performed projected crystal orbital Hamilton population (pCOHP) analysis using the LOBSTER[63,64] code. This method, applied to the VASP-calculated wavefunctions with *spd* basis for Mn and *sp* basis for N, quantifies orbital-pair bonding strength via the integrated pCOHP (IpCOHP). It allowed us to directly compare the strain evolution of two competing interactions: the Mn3c-N orbital hybridization and the direct Mn3c-Mn3c orbital overlap.

**A. Strain Modulation of Scalar Spin Chirality**

The crystal structure of $Mn_4N$ is illustrated in Fig. 1(a). This compound crystallizes in the cubic space group $Pm\overline{3}m$ (No. 221). The Mn atoms at the corner sites are denoted as Mn1a, those at the face-center sites as Mn3c, and the body-center site is occupied by N atoms. Both collinear and noncoplanar ferrimagnetic configurations have been considered in our calculations. For the collinear ferrimagnetic configuration, two spin alignment models have been reported in the literature: in one model, the spins are aligned along the [111] direction[46], while in the other, they are aligned along the [001] direction[49,65]. We therefore first calculated the magnetocrystalline anisotropy of $Mn_4N$ and found that the easy axis lies along the [111] direction. Fig. 1(a) shows the results, the spins within the Mn1a and Mn3c sublattices are parallel and aligned along the [111] direction, whereas the spins between the Mn1a and Mn3c sublattices are antiparallel. For the noncoplanar ferrimagnetic configuration, we also calculated its magnetocrystalline anisotropy. The results are shown in Fig. 1(b), where the spins of the Mn1a sites are found to align along the [111] direction, in agreement with previous reports in the literature [46,49].

To characterize the noncoplanar spin arrangement, we introduce the angle $\theta$ between the spins of the Mn1a site and the Mn3c site, which is directly extracted from the converged local magnetic moments according to

$$\theta = \arccos\left(\frac{\boldsymbol{S}_{1a} \cdot \boldsymbol{S}_{3c}}{|\boldsymbol{S}_{1a}||\boldsymbol{S}_{3c}|}\right) \tag{5}$$

Where $\boldsymbol{S}_{1a}$ and $\boldsymbol{S}_{3c}$ denote the local magnetic moments on the Mn1a and Mn3c sites, respectively. According to previous reports in the literature [44,46,49], the values of $\theta$ are 180° and 110° for the collinear (Fig. 1(a)) and noncoplanar (Fig. 1(b)) ferrimagnetic configurations, respectively.

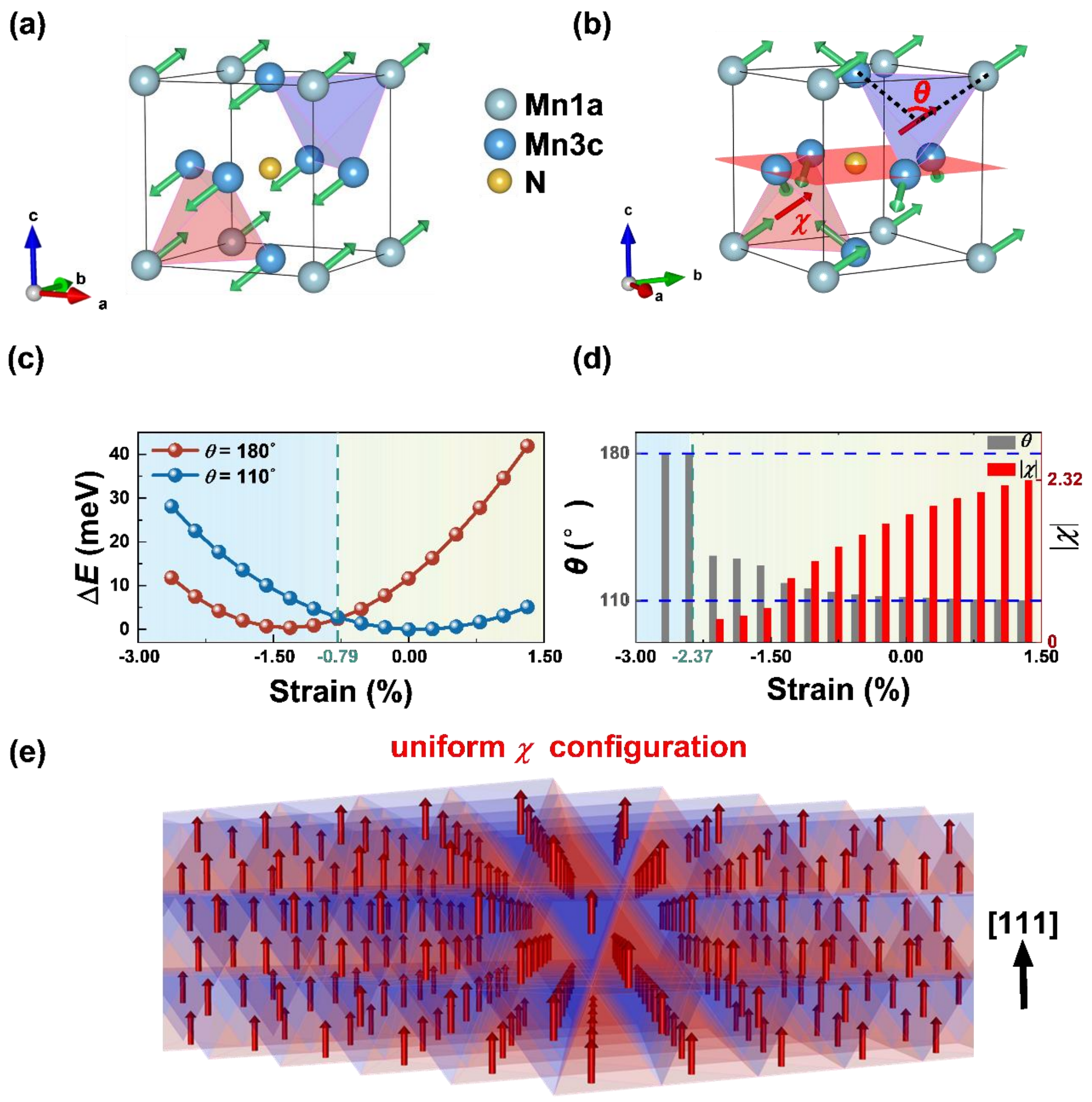

Figure 1. Collinear (a) and noncoplanar (b) magnetic configurations for cubic $Mn_4N$. $\theta$ denotes the angle between the Mn1a and Mn3c spins. The red and blue tetrahedra represent the two types of magnetic tetrahedra in the lattice. The green arrows indicate the spin directions, while the red arrows indicate the chirality-order vectors $\chi$ of the corresponding tetrahedra. (c) Strain dependence of the energy difference $\Delta E$ for the constrained magnetic configurations with $\theta = 110°$ and $\theta = 180°$. (d) Strain dependence of the equilibrium canting angle $\theta$ and the magnitude of the chirality-order vector $|\chi|$ obtained from unconstrained self-consistent calculations. (e) Uniform chirality configuration obtained from the fully unconstrained noncollinear supercell calculation.

To investigate the effect of strain on spin configuration, a zero-strain point was first defined, corresponding to the equilibrium lattice parameter $a = b = c = 3.80$ Å of the noncoplanar $Mn_4N$, which is consistent with the value reported in Ref. [46]. Fig. 1(c) shows the strain dependence of the energy difference $\Delta E$ relative to the zero-strain point, where the spin directions are constrained to $\theta = 110°$ and $\theta = 180°$ for the noncoplanar and collinear magnetic configurations, respectively. At zero strain, the noncoplanar configuration is energetically favored over the collinear configuration by approximately 10 meV/atom, which is in good agreement with the theoretical value of 8.5 meV/atom reported in Ref. [46]. Under tensile (positive) strain, the $\theta = 110°$ configuration remains energetically lower than the $\theta = 180°$ configuration, indicating that the constrained noncoplanar magnetic configuration is favored. Under compressive (negative) strain, this noncoplanar configuration remains energetically favorable down to a critical strain of -0.79%. Beyond this point, the collinear ferrimagnetic state becomes progressively more stable. The above trend is consistent with that reported in Ref. [46]. Therefore, isotropic strain can induce a transition between the noncoplanar and collinear ferrimagnetic states. In addition, previous studies have shown that biaxial strain can also drive the same transition.[49] To obtain a rough indication of how thermal expansion may affect the relative energetics of the competing magnetic configurations, we considered the experimentally reported thermal expansion of $Mn_4N$. The lattice constants corresponding to 300 K and 740 K are estimated to be approximately 3.85 Å and 3.89 Å, respectively. Static DFT calculations at these lattice constants yield energy differences between the noncoplanar and collinear configurations of approximately 108 meV/f.u. and 264 meV/f.u., respectively. For comparison, $k_BT$ is approximately 26 meV at 300 K and 64 meV at 740 K. Over this lattice-parameter range, the noncoplanar configuration remains energetically favored in the static DFT calculations. Although the corresponding energy differences are larger than the $k_B T$ scales at 300 and 740 K, this comparison should be regarded only as a qualitative energetic reference because finite-temperature spin fluctuations and entropic effects are not included.

In the calculations presented in Fig. 1(c), the canting angle $\theta$ is constrained to fixed values of 110° and 180°. Therefore, these calculations do not necessarily correspond to the fully relaxed magnetic ground state. To determine the true magnetic ground state under each strain, additional unconstrained self-consistent calculations are required. Besides the canting angle $\theta$, the magnitude of the chirality-order vector $|\boldsymbol{\chi}|$ is used to characterize the collective chirality of the magnetic tetrahedra shown in Fig.1 (a-b). The strain dependences of $\theta$ and $|\boldsymbol{\chi}|$ for the relaxed magnetic ground state are summarized in Fig. 1(d). The unconstrained self-consistent calculations show that the critical strain for the transition from the collinear to the noncoplanar magnetic configuration is -2.37%, whereas the corresponding critical strain obtained from the constrained calculations is -0.79%. Moreover, the equilibrium canting angle $\theta$ is not fixed at 110° but decreases continuously with increasing strain, gradually approaching 110° once the strain exceeds zero strain. Consequently, $|\boldsymbol{\chi}|$ increases continuously from 0 to 2.32. These results demonstrate that isotropic strain provides a continuous tuning parameter for the magnitude of the chirality-order vector.

According to Eq. (4), the chirality-order vectors $\boldsymbol{\chi}$ associated with the red and blue tetrahedra in Fig. 1(b) are both oriented along the [111] direction, as illustrated by the solid red arrows. Their parallel alignment represents a uniform chirality configuration within the magnetic unit cell. The preferred orientation of the chirality-order vector is determined by the magnetocrystalline anisotropy arising from spin–orbit coupling (SOC). For a given <111> axis, the two time-reversal-related chirality configurations are energetically degenerate.[66] In addition, symmetry-equivalent chirality domains can occur along the other <111>-type directions. To verify the time-reversal degeneracy, we calculated the energy of the opposite-chirality configuration and confirmed that it is energetically degenerate with the configuration shown in Fig. 1(b).

Since calculations using a primitive periodic unit cell necessarily repeat the same magnetic pattern, we further examined the robustness of the uniform chirality configuration using a larger supercell. Fully unconstrained noncollinear calculations were performed, and the resulting configuration is shown in Fig. 1(e). The relaxed supercell retains a uniform chirality configuration, with the chirality-order vectors associated with the magnetic tetrahedra consistently aligned along the [111] direction. This result supports the stability of the uniform-chirality solution beyond the primitive unit cell, but does not establish the formation of a macroscopic single-domain chiral state in a real sample. Because opposite-chirality domains and symmetry-equivalent <111> domains are energetically degenerate, different chirality domains may coexist in a macroscopic sample. The selection of a specific chirality domain under experimental conditions is beyond the scope of the present study.

**B. Two Parallel Consequences**

To understand the formation mechanism of noncoplanar ferrimagnetic states in $Mn_4N$, we analyzed spin-related Hamiltonians where only the Heisenberg exchange interaction and the Dzyaloshinskii–Moriya interaction influence the $\theta$ between spins. In the centrosymmetric $Mn_4N$ system, the Dzyaloshinskii–Moriya interaction is symmetry-forbidden, the $\theta$ is governed predominantly by the Heisenberg exchange interaction, which is described by $J_{ij}\boldsymbol{S}_i\cdot\boldsymbol{S}_j$. Here, $J_{ij}$ represents the exchange coupling constant, and $\boldsymbol{S}_i$ and $\boldsymbol{S}_j$ denote the local spin at $i$ and $j$ sites. Previous first-principles studies[41] reported that the next-nearest-neighbor exchange interactions are substantially smaller than the nearest-neighbor contribution. This suggests that the nearest-neighbor interaction provides a reasonable first-order description of the strain-dependent magnetic evolution discussed here.

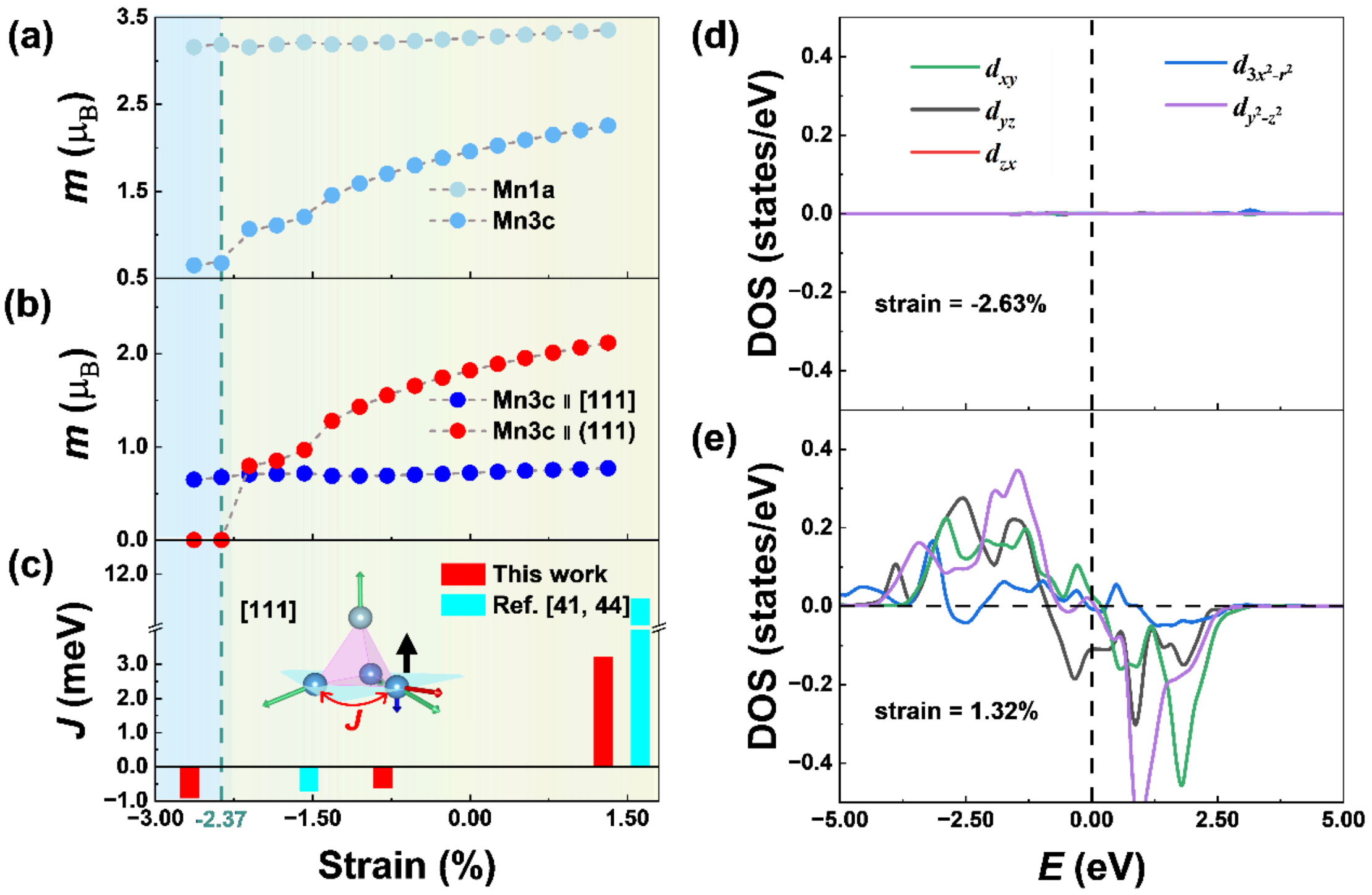

Figure 2. (a) Strain dependence of the Mn moments at Mn1a and Mn3c sites. (b) Strain evolution of the Mn3c moment components: Mn3c∥[111] (parallel to [111] crystal direction) and Mn3c∥(111) (lying within the (111) lattice planes), corresponding to the blue and red arrows illustrated in the inset of (c), respectively. (c) Strain variation of the nearest-neighbor exchange coupling constant $J$ between Mn3c sites. Projected density of states (PDOS) of the Mn3c 3$d$ orbitals onto the (111) plane at strains of -2.63% (d) and 1.32% (e).

We calculated the strain dependence of the Mn moments at the Mn1a and Mn3c sites and the nearest-neighbor Mn3c-Mn3c exchange coupling constant $J$ to quantify this interaction, as shown in Figs. 2(a-c). The Mn moment at the Mn1a site remains nearly constant at ~3.22 $\mu_B$ under strain, whereas the Mn moment at the Mn3c site, which is highly sensitive to strain, increases from 0.6 $\mu_B$ to 2.3 $\mu_B$ as strain varies from -2.63% to 1.32%. Under zero strain, our calculated value is 1.06 $\mu_B$/f.u., which is nearly the same as the neutron diffraction value (1.10 $\mu_B$/f.u.)[34].

We note that the Mn3c moments exhibit a discontinuous change at the -2.37% strain point where the system underwent a magnetic transition from noncoplanar to collinear. To further analyze the moment at Mn3c site, the total Mn moment at Mn3c site is decomposed along two directions: parallel to the [111] direction, denoted as Mn3c‖[111] (blue arrow) which is parallel to Mn1a moment; and parallel to the (111) plane, denoted as Mn3c‖(111) (red arrow) which is vertical to Mn1a moment, as shown in the inset of Fig. 2(c).

Fig. 2(b) shows the strain dependence of Mn3c‖[111] and Mn3c‖(111). It is interesting to observe that the Mn3c‖[111] component remains nearly constant at 0.7 $\mu_B$ over the whole strain region, similar to the behavior of the Mn1a moment. In contrast, the Mn3c‖(111) component emerges from zero at a critical strain of -2.37% and continuously increases with increasing strain, following a trend similar to that of Fig. 1(d). This indicates that the Mn3c‖(111) moment is a key element in constituting the noncoplanar ferrimagnetic state, i.e., the $\chi$ order. At 1.32% strain, our calculated Mn3c moment (2.31 $\mu_B$) agrees well with that (2.35 $\mu_B$) of the reference[46].

Since the Mn3c‖(111) and Mn1a moments are mutually orthogonal, the formation of $\chi$ is dictated exclusively by the exchange interactions among the Mn3c sites within the Heisenberg model. In this context, the nearest-neighbor Mn3c-Mn3c exchange coupling constant $J$ is calculated and shown in Fig. 2(c) at three representative strain points: the maximum compressive strain (-2.63%), the maximum tensile strain (1.32%), and the strain corresponding to the energy crossover point in Fig. 1(c) (-0.79%). The calculated exchange coupling constant $J$ evolves from -0.9 meV at -2.63% strain to -0.6 meV at -0.79%, and finally reverses sign to 3.2 meV at 1.32% strain. Thus, a clear strain dependence of the exchange coupling constant $J$ is observed. This evolution indicates that tensile strain weakens the ferromagnetic coupling and drives the system toward antiferromagnetic coupling between the nearest-neighbor Mn3c sites. As indicated by the cyan points in Fig. 2(c), this trend is in excellent agreement with previously reported values at comparable lattice constants[41,44].

To understand the microscopic origin of this anomalous magnetic transition, we present the projected density of states (PDOS) of the Mn3c 3$d$ orbitals onto the $(111)$ plane at the representative strains of -2.63% and 1.32%, as shown in Figs. 2(d, e). At a compressive strain of -2.63%, the in-plane PDOS contribution near the Fermi level is virtually zero. In stark contrast, under a tensile strain of 1.32%, a pronounced spin-polarized contribution strongly emerges across the $d$ orbitals. This provides direct energy-space evidence for the orbitally-driven activation of the Mn3c$\parallel$(111) moment.

Consequently, the formation of the chirality-order vectors is accompanied by two parallel phenomena: the activation of the Mn3c$\parallel$(111) moment and a sufficiently weak ferromagnetic exchange interaction (or even an antiferromagnetic exchange interaction) between the nearest-neighbor Mn3c sites.

### C. Bond Selectivity of Strain

Since $J$ is governed by the electron-orbital overlap, while the Mn3c$\parallel$(111) moment is modulated by the chemical bond of the Mn3c orbitals, it is imperative to quantify the orbital responses to strain. Charge density difference (CDD) mapping quantitatively delineates the spatial redistribution of electrons, providing a direct visual representation of these orbital responses.

In order to reflect the changes in the charge density of each component atom in $Mn_4N$, we performed the following CDD calculation:

$$\mathrm{CDD} = \rho_{\mathrm{Mn_4N}} - \rho_{\mathrm{Mn_1}} - \rho_{\mathrm{Mn_2}} - \rho_{\mathrm{Mn_3}} - \rho_{\mathrm{Mn_4}} - \rho_{\mathrm{N}} \tag{6}$$

where $\rho_{\mathrm{Mn_4N}}$ represents the charge density of the complete unit cell, $\rho_{\mathrm{Mn}_\alpha}$ $(\alpha = 1,2,3,4)$ and $\rho_{\mathrm{N}}$ correspond to the charge densities of isolated Mnα and N atoms placed at their respective crystal positions, respectively.

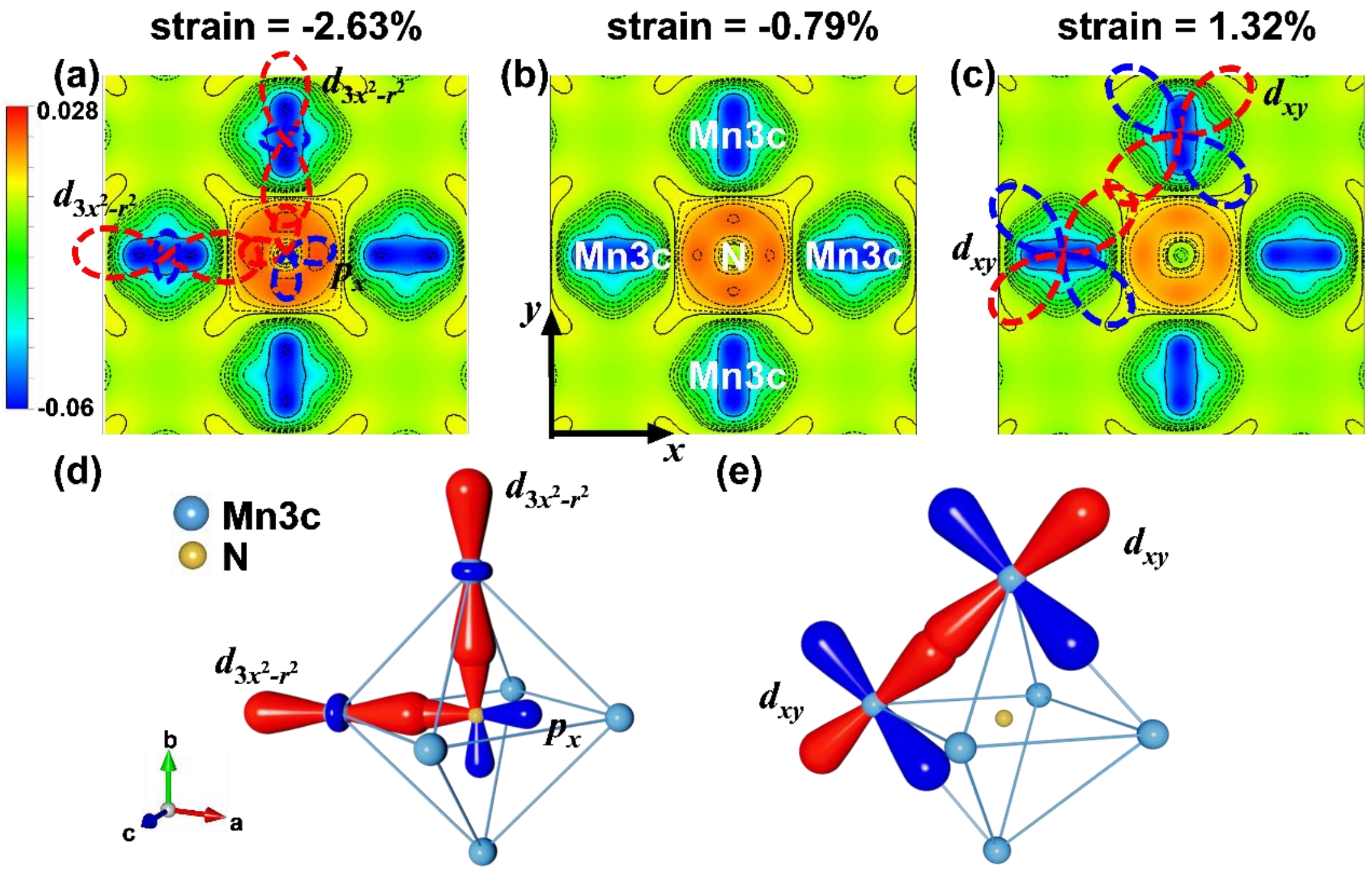

Figure 3. Charge density difference on the (001) lattice plane (corresponding to the red plane in Fig. 1(b)) under strains of -2.63% (a), -0.79% (b), and 1.32% (c). The red and blue areas represent electron accumulation and depletion, respectively. The isosurface levels are set to 0.028 e/Å$^3$(red, positive) and -0.06 e/Å$^3$ (blue, negative). Black dashed lines indicate charge contour levels. The red and blue dashed loops delineate the projections of the Mn3c and N orbitals onto the (001) plane, serving as spatial

references for orbital hybridization. Schematic representations of the two orbital bonding configurations deduced from the charge density difference maps. Schematics of the polar $\sigma$ covalent bond formed by the hybridization of the Mn3c $d_{3x^2-r^2}$ ($d_{3y^2-r^2}$) orbital and the N $p_x$ ($p_y$) orbital (d) and the nonpolar $\sigma$ covalent bond between the $d_{xy}$ orbitals of nearest-neighbor Mn3c atoms (e).

Figs. 3(a-c) show the CDD distribution on the (001) lattice plane under representative strains of -2.63%, -0.79%, and 1.32%. In these maps, the red and blue regions denote electron accumulation and depletion, respectively, while the dashed loops delineate the projections of the Mn3c and N orbitals. There are two characteristics of the distribution of CDD. First, an obvious positive CDD along $x$ and $y$ axes suddenly appears between Mn3c site and N and is very close to the N atoms as indicated by the black contour lines in the red area, and an obvious negative CDD of the Mn3c atom only along $x$ or $y$ axis also appears at the same time as shown by the black contour line in the blue area. Thus, the CDD distribution is strongly anisotropic, and meets the characteristics of the $d_{3x^2-r^2}$ and $d_{3y^2-r^2}$ orbitals of Mn atoms and the $p_x$ and $p_y$ orbitals of N atoms. That is to say, the electrons in the $d_{3x^2-r^2}$ ($d_{3y^2-r^2}$) orbital of Mn3c atoms form polar $\sigma$ covalent bonds with the electrons in the $p_x$ ($p_y$) orbital of N atoms. To intuitively illustrate this specific orbital overlap configuration, a corresponding schematic is provided in Fig. 3(d). Second, there is also a positive CDD (yellow area) between the nearest-neighbor Mn3c atoms, and its direction is 45 degrees to the coordinate axes, showing the anisotropy of the Mn3c $d_{xy}$ orbitals. This is not a characteristic of metallic bonds, but rather of covalent bonds. The nonpolar $\sigma$ covalent bond formed by the $d_{xy}$ orbitals of the Mn3c atoms is explicitly illustrated in Fig. 3(e),

which is allowed whether from the perspective of energy or orbital direction. As the tensile strain systematically increases from -2.63% to 1.32%, there are also two characteristics on the value of CDD. First, the value of CDD at the Mn3c $d_{3x^2-r^2}$ orbital and the N $p_x$ orbital overlap is significantly decreased, reflecting a severe attenuation of the polar $\sigma$ covalent bond. In stark contrast, the value of CDD between the overlapping $d_{xy}$ orbitals of the nearest-neighbor Mn3c sites remains virtually unchanged, implying that the nonpolar $\sigma$ covalent bond is essentially insensitive to the strain.

While the CDD analysis offers an intuitive spatial perspective, a rigorous quantitative evaluation of the bonding interactions is urgently needed. To this end, we employed the projected crystal orbital Hamilton population method (pCOHP).[63,64] This analytical technique mathematically extracts the specific bonding and antibonding contributions between selected atomic orbitals. A comprehensive analysis across all 36 conceivable orbital combinations confirms that the nearest-neighbor Mn3c interactions are predominantly governed by two distinct bonding pathways: the hybridization between the Mn3c $d_{3x^2-r^2}$ orbital and the N $p_x$ orbital, and the direct overlap between the Mn3c $d_{xy}$ orbital and the Mn3c $d_{xy}$ orbital. This dual-pathway bonding model is perfectly consistent with the preceding CDD analysis. The corresponding energy-resolved profiles for these two dominant orbital interactions under different strains are presented in Figs. 4(a-c). For both bonding mechanisms, the profiles remain positive below the Fermi level, signifying robust bonding states.

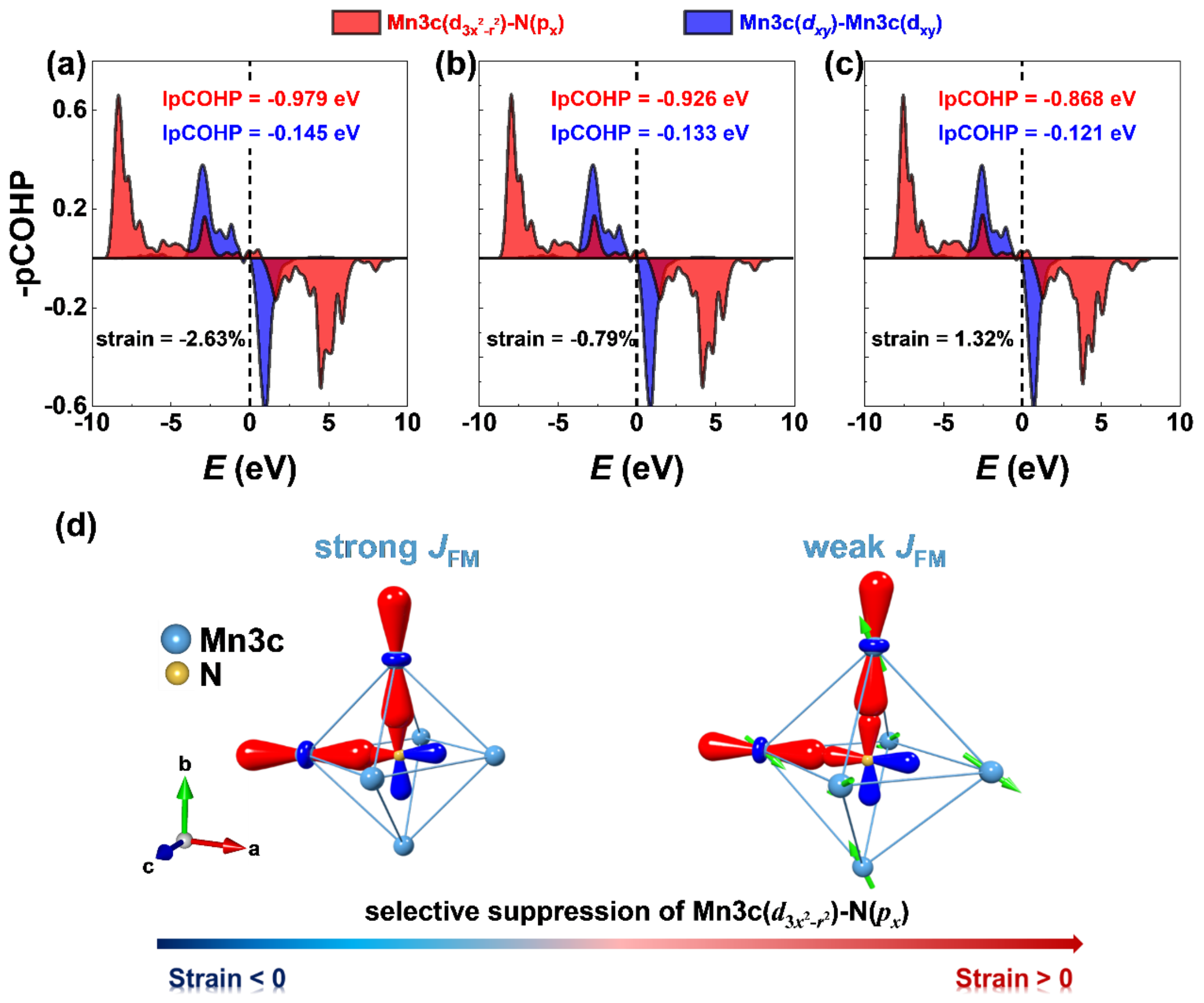

Figure 4. Projected crystal orbital Hamilton population (pCOHP) profiles for the two principal orbital interactions: the Mn3c $d_{3x^2-r^2}$ and N $p_x$ bond (red shaded area), and the Mn3c $d_{xy}$ and Mn3c $d_{xy}$ bond (blue shaded area) under strains of -2.63% (a), -0.79% (b), and 1.32% (c). Positive and negative $-\mathrm{pCOHP}$ values denote bonding and antibonding characteristics, respectively. The red and blue text in each panel gives the integrated pCOHP (IpCOHP) values up to the Fermi level (measure of bond strength) of the Mn3c-N bond and the Mn3c-Mn3c bond, respectively. (d) Strain-dependent schematic of the dual modulation of the Mn3c∥(111) moment and exchange interactions via bond-selective suppression. The left and right panels correspond to the covalent bond between the Mn3c $d_{3x^2-r^2}$ orbital and the N $p_x$ orbital and the magnetic configuration under the compressive and tensile strain conditions, respectively. The green arrows denote the activated Mn3c∥(111) moments. $J_{FM}$ represents the nitrogen-mediated ferromagnetic superexchange interaction of the nearest-neighbor Mn3c via the Mn3c ($d_{3x^2-r^2}$)-N ($p_x$) covalent bond.

The integrated projected crystal orbital Hamilton population (IpCOHP) parameter provides a measure of the bonding strength between two atoms.[63,64] The more negative the parameter, the stronger the bonding interaction. We have calculated this IpCOHP up to the Fermi level to understand how changes in strains affect the two distinct bonding pathways. Specifically, the IpCOHP value for the hybridized Mn3c $d_{3x^2-r^2}$ orbital and N $p_x$ orbital (the red text in each panel) progressively diminishes from -0.979 eV at -2.63% strain to -0.925 eV at -0.79%, and ultimately drops to -0.868 eV at 1.32% strain, with a total bond strength reduction of 0.111 eV. This systematic numerical reduction unequivocally demonstrates the pronounced suppression of this bonding channel, quantitatively confirming that the polar $\sigma$ covalent bond is severely compromised under tensile strain. Conversely, the IpCOHP values for the overlapping $d_{xy}$ orbitals between the nearest-neighbor Mn3c sites decreased by only 0.024 eV throughout the strain range, remaining essentially unchanged. This quantitative stability is in perfect agreement with the strain insensitivity intuitively inferred from the CDD analysis. Quantitative IpCOHP analyses reveal a striking bond-selective response to the strain: strain strongly affects only the $d_{3x^2-r^2}$ orbital of Mn3c, while the $d_{xy}$ orbital of Mn3c is largely unaffected.

Regarding the magnetic moment, the polar $\sigma$ covalent bond formed by the Mn3c $d_{3x^2-r^2}$ orbital and N $p_x$ orbital induces covalent spin-pairing, thereby suppressing the local magnetic moment.[67] Under compressive strain in the left panel of Fig. 4(d), the polar $\sigma$ bond suppresses the local magnetic moment. This state is corroborated by the virtually zero in-plane density of states near the Fermi level shown in Fig. 2(d). As the tensile strain increases, the strain exhibits bond selectivity. It selectively weakens the polar $\sigma$ covalent bond, which mitigates covalent spin-pairing and activates the Mn3c∥(111) moment indicated by the green arrows in the right panel of Fig. 4(d). This activation corresponds to the spin polarization emerging in Fig. 2(e). This evolution corresponds to Fig. 2(b).

Regarding the exchange interaction, the nonpolar $\sigma$ covalent bond between the $d_{xy}$ orbitals of Mn3c sites establishes a direct exchange interaction pathway, favoring an antiferromagnetic exchange interaction $J_{AFM}$,[68,69] while the polar $\sigma$ bond between the Mn3c $d_{3x^2-r^2}$ orbital and N $p_x$ orbital establishes a 90° superexchange pathway, generating a ferromagnetic exchange interaction $J_{FM}$[68,69]. Here, $J_{FM}$ and $J_{AFM}$ are introduced only as phenomenological quantities for qualitative discussion, representing the two microscopic exchange channels that contribute to the total nearest-neighbor exchange interaction $J$ shown in Fig. 2(c). The stronger the Mn3c-N (Mn3c-Mn3c) bond, the stronger the $J_{FM}$ ($J_{AFM}$). When the strain varies from -2.63% to 1.32%, the Mn3c-N bond is selectively suppressed, resulting in the weakening of $J_{FM}$, as shown in Fig. 4(d). During this process, the competition between the $J_{AFM}$ and $J_{FM}$ brings a new balance, leading to the exchange interaction $J$ in Fig. 2(c) to change from ferromagnetic (negative) to antiferromagnetic (positive). The competition between the $J_{AFM}$ and $J_{FM}$ provides a physically intuitive picture for understanding the strain-dependent evolution of the calculated exchange parameter $J$.

The strain-induced weakening of the Mn3c-N covalent bond simultaneously leads to two parallel consequences: activation of the Mn3c in-plane magnetic moment and suppression of the N-mediated ferromagnetic superexchange between neighboring

Mn3c atoms. The combined effect of these two parallel processes drives the evolution toward the noncoplanar magnetic state and enhances the associated chirality.

Finally, we discuss the possible experimental realization of the predicted strain-enhanced chirality in $Mn_4N$. Experimentally, $Mn_4N$ is commonly synthesized as epitaxial thin films, where the lattice mismatch with the substrate generates biaxial rather than isotropic strain. To evaluate the applicability of our mechanism under realistic experimental conditions, we performed additional first-principles calculations for $Mn_4N$ epitaxially grown on tensile-strained $SrTiO_3$ (111) and MgO (111) substrates. For $Mn_4N$ grown on the $SrTiO_3$ (111) substrate, the energy difference between the collinear and noncoplanar magnetic configurations, defined as $\Delta E = E_{\mathrm{collinear}}-E_{\mathrm{noncoplanar}}$, is 19.7 meV/atom, indicating that the noncoplanar configuration is lower in energy. The corresponding $|\chi|$ is 3.71. For the MgO (111) substrate, the corresponding values are $\Delta E$ = 18.9 meV/atom and $|\chi|$ = 3.80. In both cases, the noncoplanar magnetic configuration is considerably more stable than that of the unstrained crystal ($\Delta E$ = 10 meV/atom), while $|\chi|$ is significantly enhanced. The results therefore suggest that tensile epitaxy on (111)-oriented substrates provides a promising route for stabilizing the noncoplanar magnetic state and enhancing scalar spin chirality in $Mn_4N$ thin films.

Given the established connection between scalar spin chirality and Berry-phase-driven transport phenomena discussed in the Introduction, we expect unique transport features to emerge in such films, including topological Hall effect (THE)[1-4], anomalous Hall effect (AHE)[5-8], and even in-plane anomalous Hall effect (IPAHE)[70].

# Conclusion

In summary, we have systematically investigated the evolution of the magnetic ground state and the associated chirality of the magnetic tetrahedra in $Mn_4N$ under lattice strain using first-principles calculations. Tensile strain continuously enhances the noncoplanarity of the magnetic structure and correspondingly increases the magnitude of the chirality-order vector $|\chi|$. We demonstrate that the stabilization of the noncoplanar magnetic state and the associated chirality arises from two parallel consequences: activation of the Mn3c in-plane magnetic moment and weakening of the ferromagnetic exchange interaction between nearest-neighbor Mn3c sites. Orbital-resolved electronic structure analyses, including charge density difference and projected crystal orbital Hamilton population analyses, reveal that tensile strain selectively suppresses the polar Mn3c-N covalent bond. This bond-selective electronic response simultaneously drives the two parallel consequences and stabilizes the noncoplanar magnetic state together with the associated scalar spin chirality. We propose that this bond-centric computational framework may be applicable to engineering scalar spin chirality in other ferrimagnetic systems. Our findings provide a microscopic strategy for controlling chiral magnetic states through lattice engineering.

**Data Availability**

All data generated or analyzed during this study are included in this published article.

**Acknowledgement**

The authors gratefully acknowledge the financial support provided by the National Natural Science Foundation of China (Grant No.~51971087), the "333 Talent Project" of Hebei Province (Grant No.~C20231105), the Basic Research Project of Shijiazhuang Municipal Universities in Hebei Province (Grant No.~241790617A), the Central Guidance on Local Science and Technology Development Fund of Hebei Province (Grant No.~236Z7606G), and the Science Foundation of Hebei Normal University (Grant No.~L2024B08). These funding sources have been instrumental in facilitating the completion of this research.

**Author Contributions**

X. L. performed the first-principles calculations, analyzed the data, and wrote the original draft of the manuscript. L. M. conceived the idea, supervised the project, acquired the funding, and substantially revised the manuscript. M. Z. and S. N. assisted with the structural modeling and DFT calculations. Yu. L. and Ya. L. contributed to the data visualization and the analysis of magnetic exchange interactions. J. Z., Y. Z., and F. M. participated in the rigorous discussions regarding the bond-selective mechanism and scalar spin chirality. D. Z., G. L., C. Z., and D. H. provided technical support for the electronic structure analyses and contributed to the theoretical interpretation. All authors have read and approved the final manuscript.

**Competing interests**

All authors declare no financial or non-financial competing interests.